# Electrically function switchable magnetic domain-wall memory

Yu Sheng[1], Weiyang Wang[1, 2], Yongcheng Deng[1], Yang Ji[1, 2], Houzhi Zheng[1, 2],

Kaiyou Wang[1, 2, 3*]

1 State Key Laboratory of Superlattices and Microstructures, Institute of Semiconductors, Chinese Academy of Sciences, Beijing, 100083, China.

2 College of Materials Science and Opto-Electronic Technology, University of Chinese Academy of Sciences, Beijing, 100049, China.

3 Beijing Academy of Quantum Information Sciences, Beijing, 100193, China

*Corresponding author. Email: kywang@semi.ac.cn

**ABSTRACT**

More-versatile memory is strongly desired for end-users to protect their information in the information era. In particular, bit-level switchable memory, from rewritable to read-only function, allows end-users to prevent any important data from being tampered with. However, no such switchable memory has been reported. We demonstrated the rewritable function can be converted into a read-only function by a sufficiently large current pulse in a U-shaped domain-wall memory composed of an asymmetric Pt/Co/Ru/AlO$_x$ heterostructure with strong Dzyaloshinskii-Moriya interaction. Wafer-scale switchable magnetic domain-wall memory arrays on 4-inch Si/SiO$_2$ substrate were designed and fabricated. Furthermore, we confirmed the information can be stored in rewritable or read-only state at bit-level according to the security-needs of end-users. Our work not only provides a solution for personal confidential data, but also paves the way for developing multi-functional spintronic devices.

INTRODUCTION

With the unprecedented expansion of the information technology, a large amount of data produced by ubiquitous terminals calls for more-versatile memory not only with advantageous performance but also with high security, particularly demanded by end-users. CMOS-based memory, as the current mainstream memory, is the main bottleneck of computer system [1,2]. Researchers have made great efforts on versatile memory with high-performance and low-power operation, such as resistance change memory[3-6], phase change memory[7–10], ferroelectric memory[11–14], and spintronic memory[15–20]. However, all these reported memories so far only have the rewritable function, and thus the stored information risks being tampered by hackers through internet, which cannot be fully resolved by encryption.

Here we report on experimental realization of wafer-scale domain-wall memory arrays combined with high performance and high security, electrically switchable from rewritable to read-only function at bit-level. In a U-shaped domain-wall memory composed of an asymmetric Pt/Co/Ru/AlO$_x$ heterostructure with strong Dzyaloshinskii-Moriya interaction, the rewritable function can be realized by field-free current-induced domain-wall reciprocating motion by spin orbit torques (SOTs). The stored information can be converted to read-only state bit by bit through domain-wall annihilation after applying a sufficiently large current. Thus, the read-only bits become immune to electric current pulses, ensuring information integrity under the threat of being exposed to hackers or impostors.

RESULTS

**Electrical operation of a U-shaped SOTs driven domain-wall memory device**

To effectively create and control the magnetic domain-wall by SOTs, a U-shaped device was designed consisting of asymmetric Pt(4 nm)/Co(0.85 nm)/Ru(1.2 nm) multilayers. An AlO$_x$(0.5 nm) capped layer was deposited on the curved region to locally weaken the perpendicular magnetic anisotropy[21] (Fig. 1a). Magnetic hysteresis loops were measured using a polar magneto-optical Kerr effect (pMOKE) microscope, with a focused laser spot positioned at the straight section, curved section and the boundary between them (Fig. 1b). The switching fields of the curved and straight sections are 25 Oe and 125 Oe, respectively, confirming the role of the AlO$_x$ capped layer.

Double-step switching of the hysteresis loop reveals the domain-wall can be pinned at the boundary of the straight and curved sections (Fig. 1b).

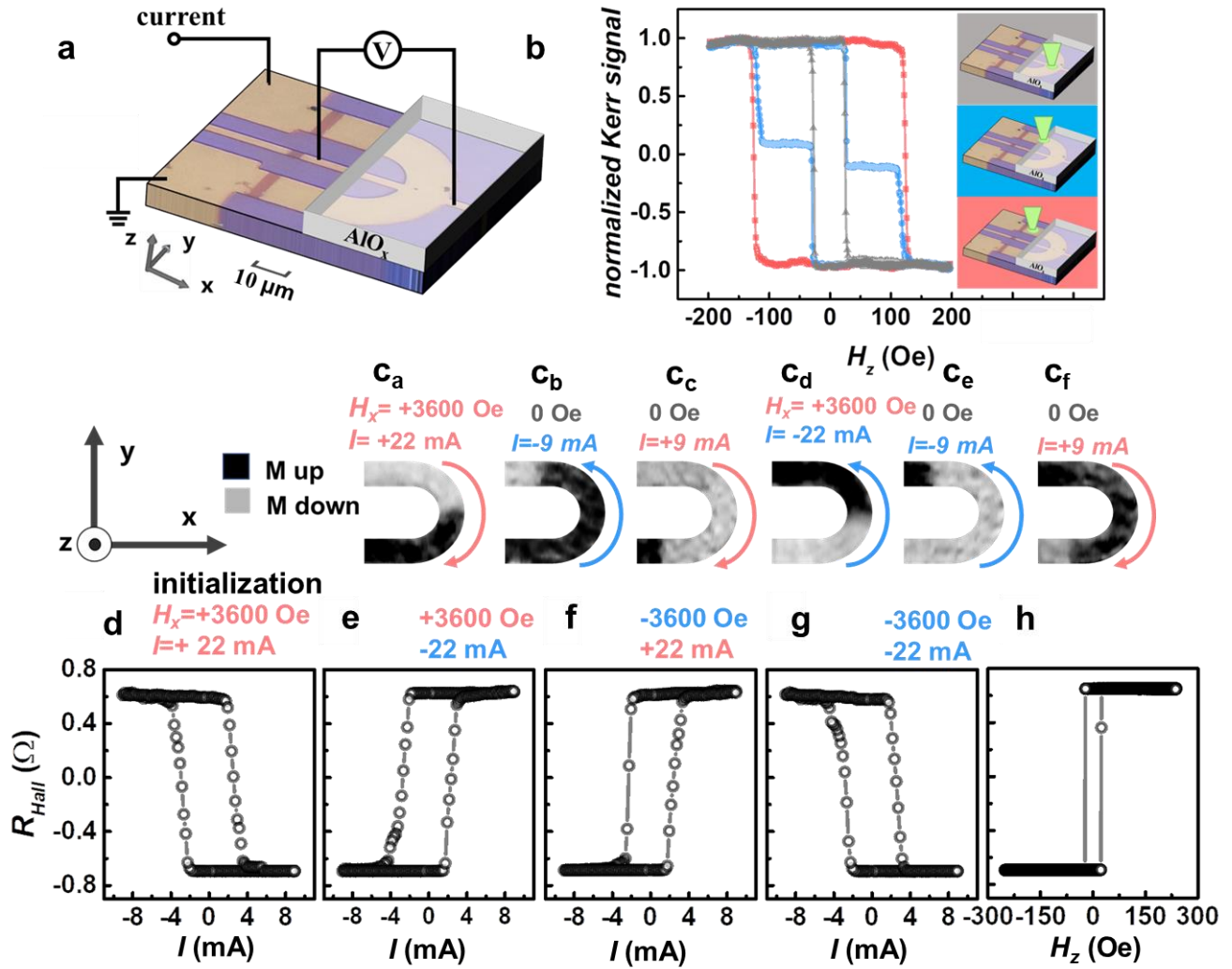

**Figure 1.** Device structure, magnetic switching properties, and magnetic field-free switching by SOTs induced domain-wall motion. (a) Optical image of the U-shaped device and anomalous Hall effect measurement configuration with the definition of x–y–z coordinates. The red-brown stripes at both ends of the U-shaped region are trenches where Co/Ru bilayer are etched away by argon ion beam. (b) Out-of-plane magnetic hysteresis loops, obtained by a polar magneto-optical Kerr effect (pMOKE) microscope, of the U-shaped device in the Pt(4 nm)/Co(0.85 nm)/Ru(1.2 nm) region (red), Pt/Co/Ru/ AlO$_x$ (0.5 nm) region (grey) and the boundary (blue), as shown in the inserts. (c$_a$-c$_c$) The pMOKE images of the magnetic domain states with initialization of $H_x$=3600 Oe and $I_{pulse}$=+22 mA (c$_a$), and the images after $I_{pulse}$=-9 mA(c$_b$)/+9 mA(c$_c$) in absence of $H_x$. (c$_d$-c$_f$) The pMOKE images with initialization of $H_x$=3600 Oe and $I_{pulse}$=-22 mA(c$_d$), and after $I_{pulse}$=-9 mA(c$_e$)/+9 mA(c$_f$) in absence of $H_x$. Dark- and light-grey regions indicate magnetization-up and -down domains, respectively. (d-g) Field-free current-induced deterministic switching loops of perpendicular magnetization in the curved section (AlO$_x$ capped) for four types of initializations:

$H_x$=±3600 Oe and $I_{pulse}$=±22 mA. (h) Out-of-plane magnetic field induced magnetization switching loop in the curved section.

Theoretically, the z-component of SOTs effective field is given by[22],

$$H_z^{SOT} = \frac{\hbar}{2eM_s t}\theta_{SH} J_x m_x \qquad (1)$$

where $\hbar$ is the reduced Plank constant, $e$ is the electron charge, $M_s$ is the saturation magnetization, $t$ is the thickness of the ferromagnetic layer, $\theta_{SH}$ is the spin Hall angle of heavy metal, $m_x$ is the x-component of magnetization orientation and $J_x$ is the x-component of charge current density. The $H_z^{SOT}$ direction is defined by both directions of $J_x$ and $m_x$. In the U-shaped device, the $J_x$ changes its direction as passing the centerline of the curved section. Therefore, the $H_z^{SOT}$ for the up- and down-regions of the device will be opposite with the same direction of $m_x$ determined by an in-plane magnetic field, leading to opposite domains formed at both sides of the centerline of the curved section, which is confirmed by the pMOKE images in Fig. 1 $c_a$ and $c_d$. The up-region prefers downward magnetization and the down-region prefers upward magnetization after initialization with $H_x$=3600 Oe and $I_x$=22 mA (Fig. 1$c_a$). But opposite magnetic domain configurations are formed after $H_x$=3600 Oe and $I_x$=-22 mA (Fig. 1$c_d$). If both the magnetic field and current directions are reversed for initializations, the same magnetic domain configurations are formed. By constructing asymmetry stacks of Pt/Co/Ru with large Dzyaloshinskii-Moriya interaction (DMI), left-hand Néel domain-wall (⊙|⊗: ↑←↓ or ⊗|⊙: ↓→↑) can be stabilized[23, 24]. Thus, the electric current can drive the domain-wall move along the current direction without the assistance of an external magnetic field.

We then investigate the field-free current-induced magnetization switching with the domain-wall motion using both anomalous Hall effect and pMOKE for different initialized magnetic domain configurations. As shown in Fig. 1d, after the initialization with $H_x$=3600 Oe and $I_{pulse}$=22 mA, a clockwise current-induced switching loop at zero field was observed. The positive current favors the domain-wall motion from up-region to down-region, while the negative current favors opposite domain-wall motion, resulting the magnetization downward. However, after the initialization with $H_x$=3600 Oe and $I_x$=-22 mA, the anti-clockwise current-induced magnetization switching were observed (Fig. 1e). As expected, for the same initialization of the magnetic domain configurations, the device shows the same deterministic current-induced magnetization switching (Fig. 1d&f and Fig. 1e&g). The switching magnitudes of $R_{Hall}$ after initializations of $H_x$=3600 Oe

& $I_x$=22 mA, $H_x$=3600 Oe & $I_x$=-22 mA, $H_x$=-3600 Oe & $I_x$=22 mA and $H_x$=-3600 Oe & $I_x$=-22 mA are 1.30 Ω, 1.31 Ω, 1.31 Ω and 1.30 Ω, respectively, which are very close to the switching amplitude of 1.34 Ω (~97%) by the magnetic field (Fig. 1h). The critical switching current densities for these four initialized configurations are among (2.5±0.2)×10$^6$ A/cm$^2$, indicating low power consumption in applications. In contrast, the referenced symmetric Pt/Co/Pt stacking U-shaped devices (Fig. S1) has much smaller switching amplitude (12%-35%) and the current-induced switching direction is un-corelated to the initialization [25,26], indicating the importance of the large DMI for SOTs based domain-wall devices.

**Narrow resistance distribution and high reproducibility**

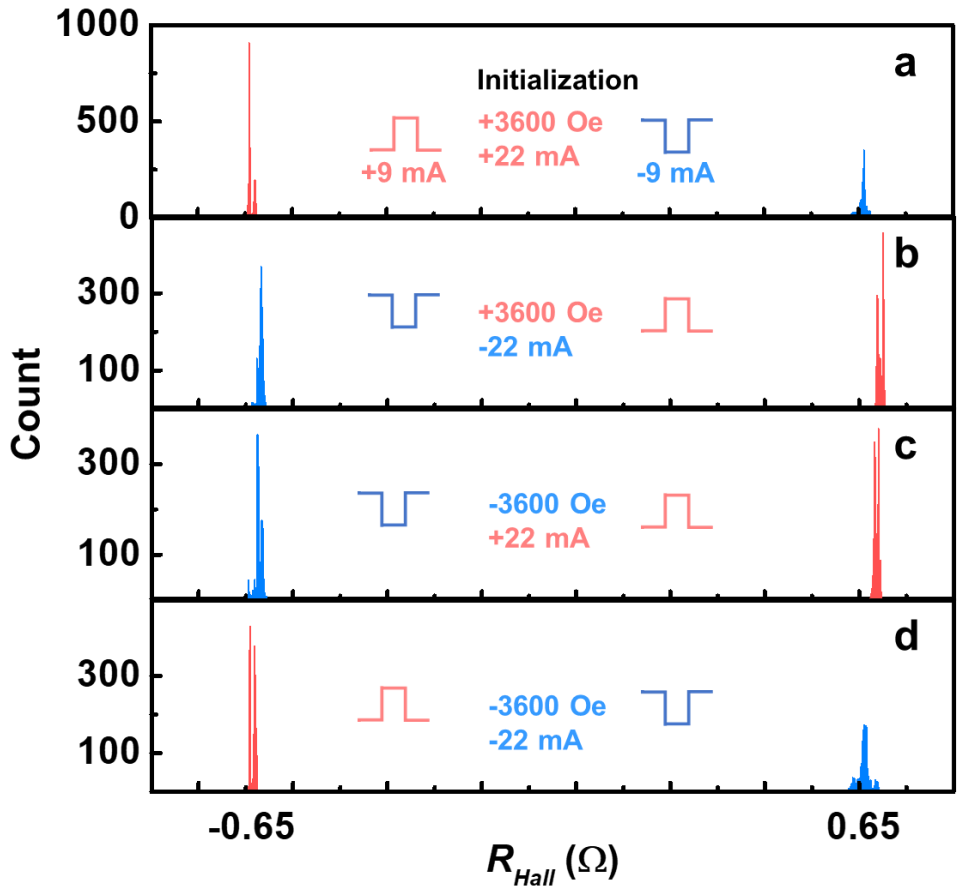

**Figure 2.** Narrow resistance distribution under alternately positive and negative current pulses. (a-d) Histograms of resistance distribution with bin size of 0.0015 Ω under alternately current pulses for all types of initializations: $H_x$=±3600 Oe and $I_{pulse}$=±22 mA. The applied pulse sequences are alternately positive and negative current pulses with amplitude of 9 mA and duration of 10 ms for 2000 cycles.

To achieve error-free rewritable domain-wall memory and robust read-out, the average change in resistance (ΔR) from the low- to high-state should be at least 12σ to have working memories with bit counts of Mb or more[27], where the standard deviation $\sigma = \sqrt{\sum_{i=1}^{n}(R_i - R_{ave})^2/n}$, $R_{ave}$ is the averaged anomalous Hall resistance $R_i$ of each state, and n=2000 is the total number of pulse cycles. For all the four types of the initializations, we realize the rewritable function with high reproducibility using alternately current pulses (Fig. 2a-d). The ratio of ΔR/σ for each type of initialization, $H_x$=3600 Oe & $I_x$=22 mA, $H_x$=3600 Oe & $I_x$=-22 mA, $H_x$=-3600 Oe & $I_x$=22 mA and $H_x$=-3600 Oe & $I_x$=-22 mA are 177, 240, 223 and 146 respectively, proving that the device shows excellent error-free switching and read-out ability. A wide current-induced switching window up to 10 mA with ΔR/σ higher than 12 is obtained for initialization of $H_x$=3600 Oe & $I_x$=22 mA (Fig. S2), ensuring stable switching for applications.

**Function switchable in a domain-wall memory device**

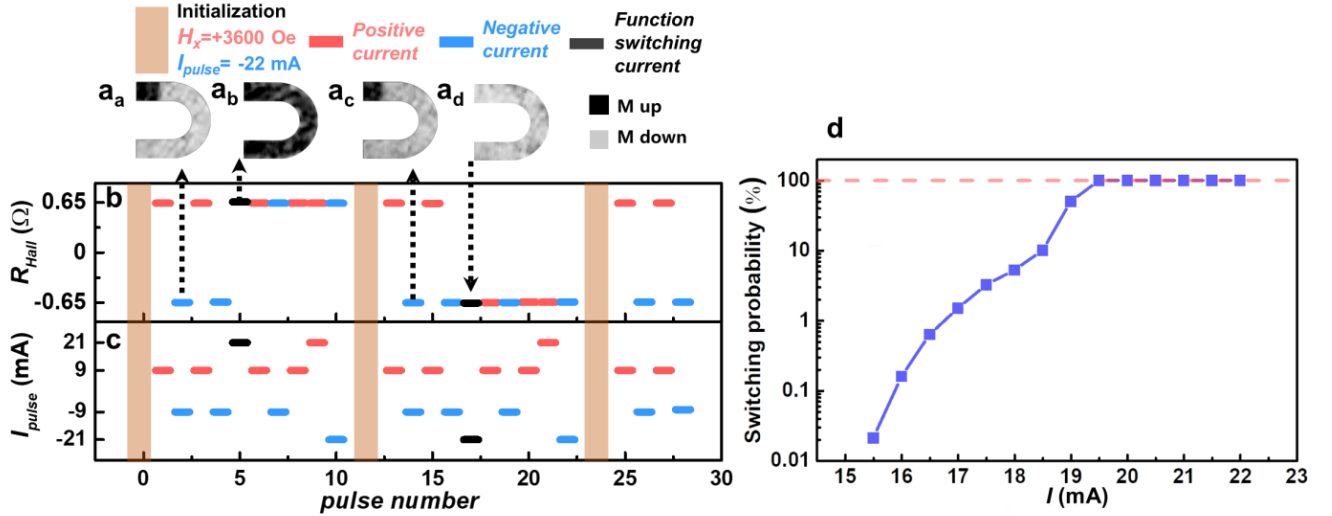

**Figure 3.** Function switching of the domain-wall device. (a-c) The measured magnetic domain-wall configurations ($a_a$-$a_d$) and $R_{Hall}$ (b) after each current pulse (c). Red (blue)-line segments indicate the positive (negative) current pulses and the corresponding $R_{Hall}$ after the current pulse, and the black-line segments indicate the current pulse for function switching from rewritable to read-only and the corresponding $R_{Hall}$. The orange area represents initialization with $H_x$=3600 Oe and $I_x$=-22mA. Inserts show pMOKE images of the corresponding magnetic states indicated by the dashed arrows ($a_a$-$a_d$). The pulse durations are all 10 ms. (d) The switching probability from rewriteable to read-only function as a function of the current pulse amplitude.

To convert the device from rewritable to read-only function, a large current pulse is needed to annihilate the domain-wall. After initialization of $H_x$=3600 Oe and $I_x$=-22 mA (Fig. 3), the +/-9 mA current pulses can drive the domain-wall back and forth to realize the rewritable function at zero magnetic field. Then, a large current pulse of +21 mA is injected into the device, and $R_{Hall}$ switched from low resistance state to high resistance state. After that, the $R_{Hall}$ remained high resistance state and cannot be changed even with very large current pulses of +/-21 mA (Fig. 3). Similarly, a -21mA current pulse leads the device to the low resistance state. Therefore, after a large current pulse applied, the device entered into the read-only function associating with the annihilation of the domain-wall, as confirmed by the pMOKE images in Fig. 3 $a_b$ and $a_d$. We further studied the dependence of switching probability from rewriteable to read-only function on the current pulse magnitude. The switching probability is less than 0.01% with current smaller than 15.5 mA, increasing to 100% with current of 19.5 mA (Fig. 3d and Fig. S4).

Interestingly, with the assistance of an in-plane magnetic field, the device in read-only function was electrically re-converted to rewritable function along with the regeneration of the domain-wall (Fig. 3c).

**Demonstration of wafer-scale function switchable domain-wall memory arrays**

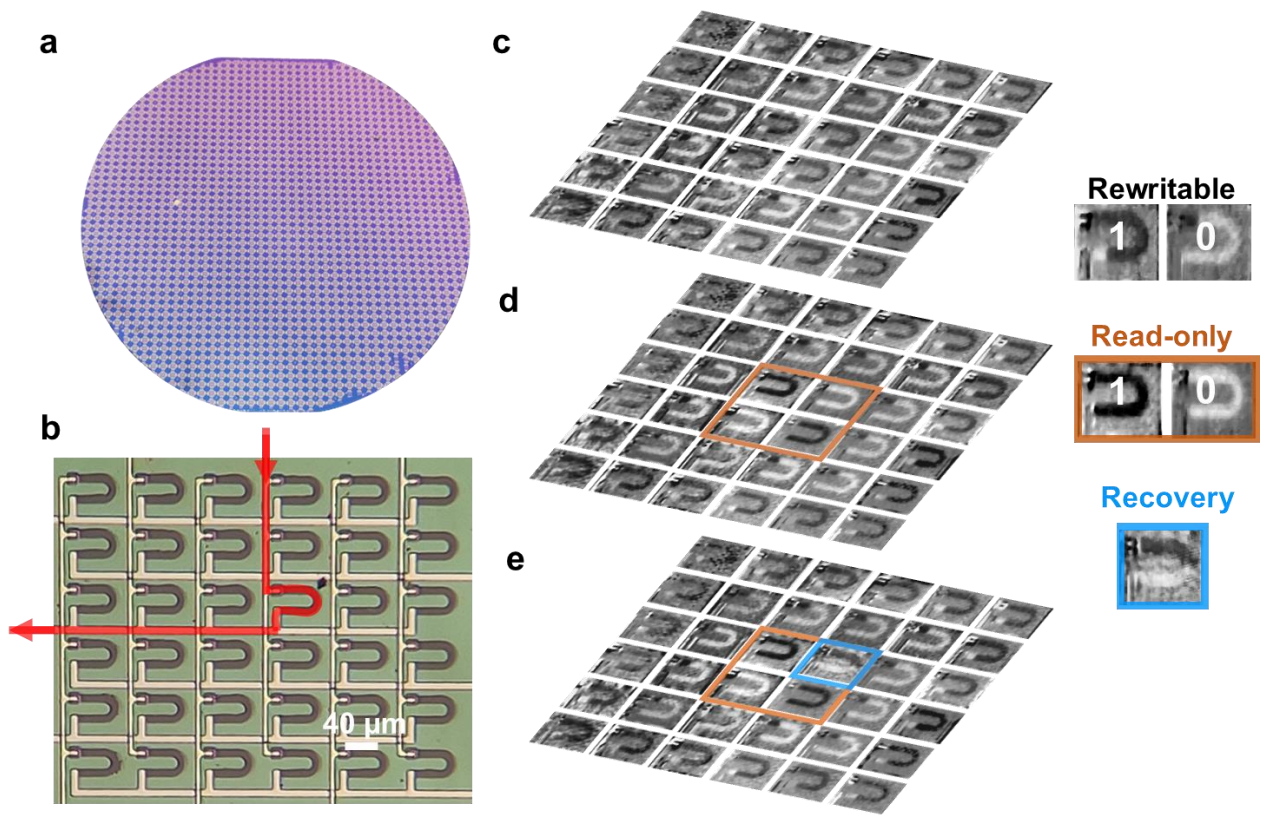

**Figure 4.** Wafer-scale function switchable domain-wall memory arrays. (a) The optical image of the function switchable magnetic domain-wall memory arrays fabricated on a 4-inch silicon wafer. (b) The optical image of the zoomed-in 6×6 array of the wafer-scale devices, where the arrowed red line indicates the path to inject a current pulse to a single device. (c-e) pMOKE images of the zoomed-in 6×6 array, right column defines the pMOKE images for '0' and '1' in rewritable and read-only state, and also recovery state. (c) Under rewritable function, a selected set of data '0' and '1' were written. (d) Four bits in the center of the array were converted to read-only function of '1' ('0') by a current pulse of +(-)21 mA, as shown in the brown frame. (e) One read-only bit was reversed back to rewritable function in combination of an in-plane magnetic field of 3600 Oe and a current pulse of -22 mA, where the brown frame changed to blue frame.

We fabricated function switchable domain-wall memories on a 4-inch Si/SiO$_2$ substrate (Fig. 4a) using standard CMOS-compatible process. The optical image of the zoomed-in 6×6 array of the wafer-scale devices is shown in Fig. 4b, where each device in the array can be electrically controlled by selecting the corresponding row and column electrodes. The information in our function switchable domain-wall memory can be stored in rewritable state or read-only state at bit-level according to the demands of end-users. A set of data in rewritable function were written by current pulses of +/- 9 mA, as shown in Fig. 4c, where the bits in the outermost circle are '1' using +9 mA (magnetization downward), the bits in the middle circle are '0' using -9 mA (magnetization upward), and the 2×2 bits in the center are '1'. Then, the 2×2 bits in the center were converted into read-only function of '1' using current pulse of +21 mA and '0' using current pulse of -21 mA (Fig. 4d, brown frames). The read-only state cannot be changed solely by electrical current since it has no domain-wall (Fig. 4d), thus eliminating the possibility of tampering with data by hackers through internet. With the aid of an in-plane magnetic field, the read-only bit cell can be electrically converted back to rewritable cell without affecting other bits (Fig. 4e, blue frame). To our knowledge, such recoverability is not available with the conventional read-only memory. The wafer-scale switchable domain-wall memory not only allows end-users to store their information in rewritable or read-only state at bit-level according to their own wishes, but also can work together with the existing encryption techniques, which can further meet customized security requirements.

**CONCLUSION**

Except for being demonstrated in spin torque-driven race-track memory[28,29], magnetic domain-walls with non-volatility and high design-flexibility have attracted a lot of interests, with remarkable achievements in magnetic field-driven logic[30], SOTs-driven magnetic logic[17,28], and neuromorphic computing[31,32]. The reciprocal motion of the current-driven domain-wall in this work could also be utilized for spin logic and neuromorphic computing. Therefore, in addition to securing stored information for end-users, CMOS-compatible switchable domain-wall devices could pave the way for developing novel computing and logic-in memory devices.

## METHODS

**Thin Film Preparation**

The films were deposited on Si/SiO$_2$ substrate by magnetron sputtering at room temperature. DC magnetron sputtering was used to deposit the Ta, Pt, Co, Ru and AlO$_x$ layers. The base pressure of the chamber was less than $2.0 \times 10^{-6}$ Pa. The pressure of the chamber was $1.06 \times 10^{-1}$ Pa during deposition. The deposition rates for Ta, Pt, Co, Ru and AlO$_x$ layers were controlled to be 0.022, 0.0240, 0.0124, 0.017 and 0.002 nm/s, respectively.

**The Device Fabrication**

For a single device, two steps of UV lithography and magnetron sputtering deposition were used. First, the Ta (1 nm)/Pt (4 nm)/Co (0.85 nm)/Ru (1.2 nm) film was patterned into U-shaped devices with channel width of 16 μm and inner diameter of 30 μm. Then, using photolithography engraving and magnetron sputtering deposition, AlO$_x$ layer with thickness of 0.5 nm was grown in the curved section. Finally, argon ion beam etching is used to make trenches with width of 5 μm in both straight sections, where Co/Ru bilayers were precisely etched away.

For the wafer-scale domain-wall memory arrays, every U-shaped device has channel width of 10 μm and inner diameter of 20 μm. Additional process of lift-off is needed to make the row electrodes (Ta(10 nm)/Au (50 nm)) and column electrodes (Ta(10 nm)/Au (150 nm)). To prevent the electrical contact between the two layers of electrodes, a layer of SiO$_x$ with thickness of 100 nm is deposited by PECVD (plasma-enhanced chemical vapor deposition) to separate them.

**Measurement and Characterization**

The current-induced magnetization switching and anomalous Hall effect measurements were carried out at room temperature with a Keithley 2602B as the current source and Keithley 2182 as the nanovoltmeter. The Kerr imaging measurements were carried out using a NanoMOKE3 microscope.

**SUPPLEMENTARY DATA**

Supplementary data are available at NSR online.

**FUNDING**

This work was supported by the Beijing Natural Science Foundation Key Program (Grant No. Z190007), by the NSFC (Grant Nos. 11474272, 61774144 and 12104449), by the Chinese Academy of Sciences (Grant Nos. QYZDY-SSW-JSC020, XDPB44000000, and XDB28000000).

**AUTHOR CONTRIBUTIONS**

K.W. conceived the work. Y.S. and W. W. deposited the thin films. Y. S. fabricated the devices and performed the SOTs-driven DW measurements. Y.S. and K. W. analyzed the data and wrote the manuscript. All authors discussed the results and commented on the manuscript.

*Conflict of interest statement.* None declared.

# SUPPLEMENTARY DATA

## Electrically switchable magnetic domain-wall memory

Yu Sheng[1], Weiyang Wang[1, 2], Yongcheng Deng[1], Yang Ji[1, 2], Houzhi Zheng[1, 2], Kaiyou Wang[1, 2, 3*]

1 State Key Laboratory of Superlattices and Microstructures, Institute of Semiconductors, Chinese Academy of Sciences, Beijing, China.

2 College of Materials Science and Opto-Electronic Technology, University of Chinese Academy of Sciences, Beijing, China.

3 Beijing Academy of Quantum Information Sciences, Beijing 100193, China

[*]Corresponding e-mail: kywang@semi.ac.cn

**Contents**

S1. Magnetic properties and current-induced magnetization switching of the symmetric U-shaped Pt(4 nm)/Co(0.5 nm)/Pt(1.2 nm) device

S2. Endurance measurements of Pt(4 nm)/Co(0.5 nm)/Ru(1.2 nm) U-shaped devices

S3. Switching probability of the function switchable device from rewritable to read-only function

## S1. Magnetic properties and current-induced magnetization switching of the symmetric U-shaped Pt(4 nm)/Co(0.5 nm)/Pt(1.2 nm) device

In order to reveal the importance of Dzyaloshinskii−Moriya interaction (DMI) in field-free current-induced magnetization switching through domain wall motion, we fabricated the U-shaped device as reference based on Pt/Co/Pt stacks, where the DMI is very small due to the symmetric structure[1].

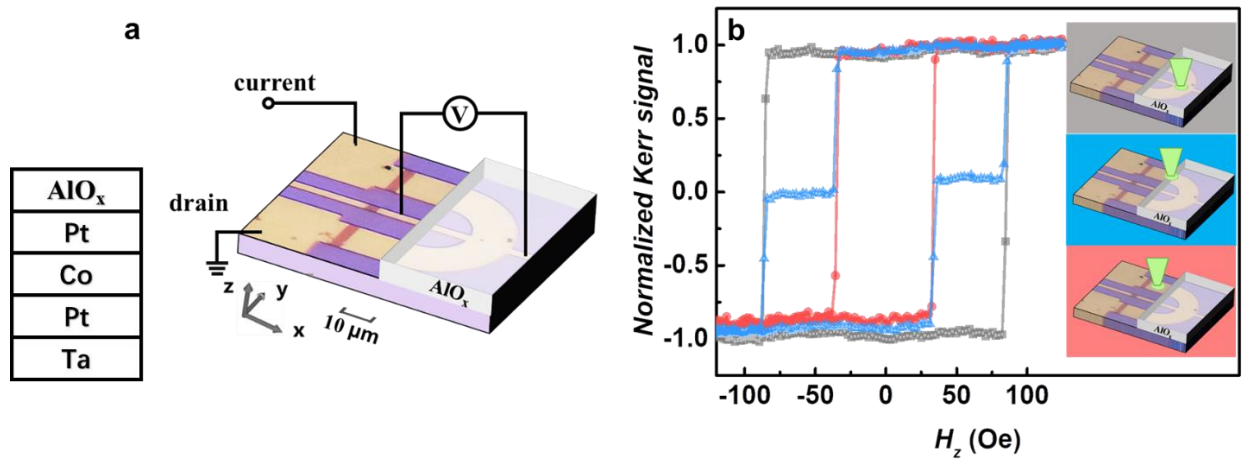

**Figure S1.** (a) Optical image of the U-shaped device and anomalous Hall effect measurement configuration for symmetric Pt/Co/Pt stacks with definition of x–y–z coordinates. The Co/Ru bilayer at both ends of the U-shaped region (red-brown stripes) are etched away by argon ion beam. (b) Out-of-plane hysteresis loops, obtained by a polar magneto-optical Kerr effect (pMOKE) microscope, of the U-shaped device in the Pt/Co/Pt region (red), Pt/Co/Ru/AlO$_x$ region (grey) and at the boundary (blue), as the inserts shown. The green cone in the inserting image represents the detecting laser from the pMOKE microscope.

As shown in Fig. S1, except for using the symmetric Pt/Co/Pt stacks, the U-shaped device structure is the same to that in Fig. 1a. The AlO$_x$ capped layer was deposited on the curved region to locally weak the perpendicular magnetic anisotropy, creating a potential step at the boundary between the two regions (Fig. S1a). Magnetic hysteresis loops were measured using a polar magneto-optical Kerr effect (pMOKE) microscope, with a focused laser spot of 2 μm positioned at the straight section, curved section and the boundary between them (Fig. S1b). The switching fields in curved section and straight section are 35 Oe and 86 Oe, respectively, confirming the role of AlO$_x$ capped layer. Double-step switching of the hysteresis loop at the boundary reveals the domain-wall can be pinned at the boundary (Fig. S1b).

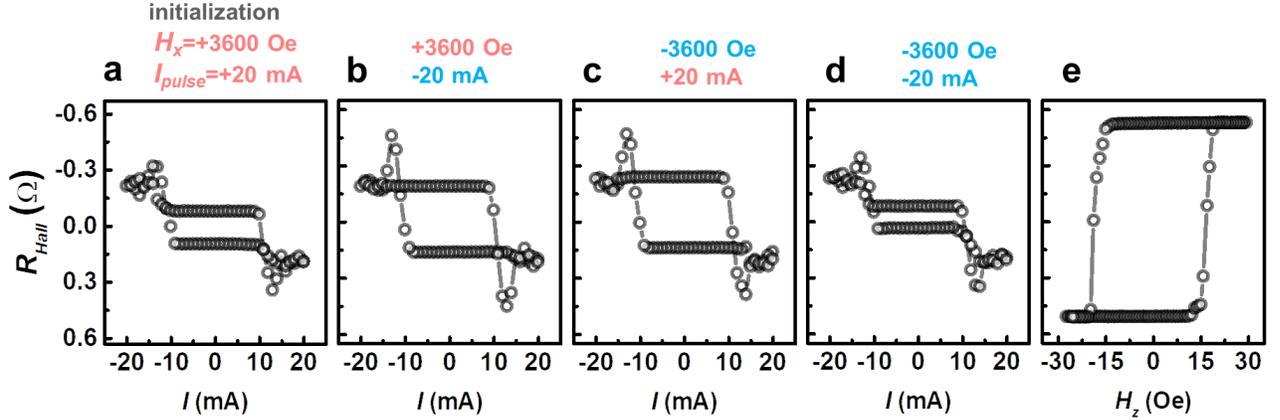

**Figure S2.** (a-d) Spin-orbit torques induced magnetization switching of perpendicular magnetization in the curved section (AlO$_x$ capped) for four types of initializations: $H_x=\pm 3600$ Oe, $I_{pulse}=\pm 20$ mA. The applied pulse sequences for (a)-(d) are pulse strings with each duration of 10 ms and scanning magnitude between the maximum $I_{pulse}=\pm 20$ mA, and all $R_{Hall}$ data point are measured 100 ms after the application of each pulse under a detecting current of 0.1 mA. (e) Out-of-plane magnetic field induced switching loop of the perpendicular magnetization in the curved section. All the data are measured for the device with symmetric Pt/Co/Pt structure.

We then investigated the current-driven domain-wall motion in the symmetric U-shaped device. In Fig. S2a-d, four types of initializations with $H_x=\pm 3600$ Oe and $I_{pulse}=\pm 20$ mA were performed to create opposite domain in two straight sections, and then we measured the anomalous Hall resistance vs. the electrical current pulses after each initialization. Similar clockwise-like switching loops were observed in response to current pulse for all four types of initializations, where the switching magnitudes of $R_{Hall}$ after initializations of $H_x=3600$ Oe & $I_x=22$ mA, $H_x=3600$ Oe & $I_x=-22$ mA, $H_x=-3600$ Oe & $I_x=22$ mA and $H_x=-3600$ Oe & $I_x=-22$ mA are 0.17 Ω, 0.35 Ω, 0.38 Ω and 0.11 Ω, respectively, with the switching amplitudes of $R_{Hall}$ being much smaller than the switching amplitude caused by perpendicular magnetic field (1.04 Ω) (Fig. S2e). In contrast to the asymmetric Pt/Co/Ru device, the symmetric Pt/Co/Pt with small DMI prohibits the current-induced domain-wall motion reciprocally[25,26]. Thus, the rewriteable function cannot be realized in this symmetric structure device, confirming the importance of the DMI for the SOT driven domain-wall memory.

**S2. Endurance measurements of Pt(4 nm)/Co(0.5 nm)/Ru(1.2 nm) U-shaped devices**

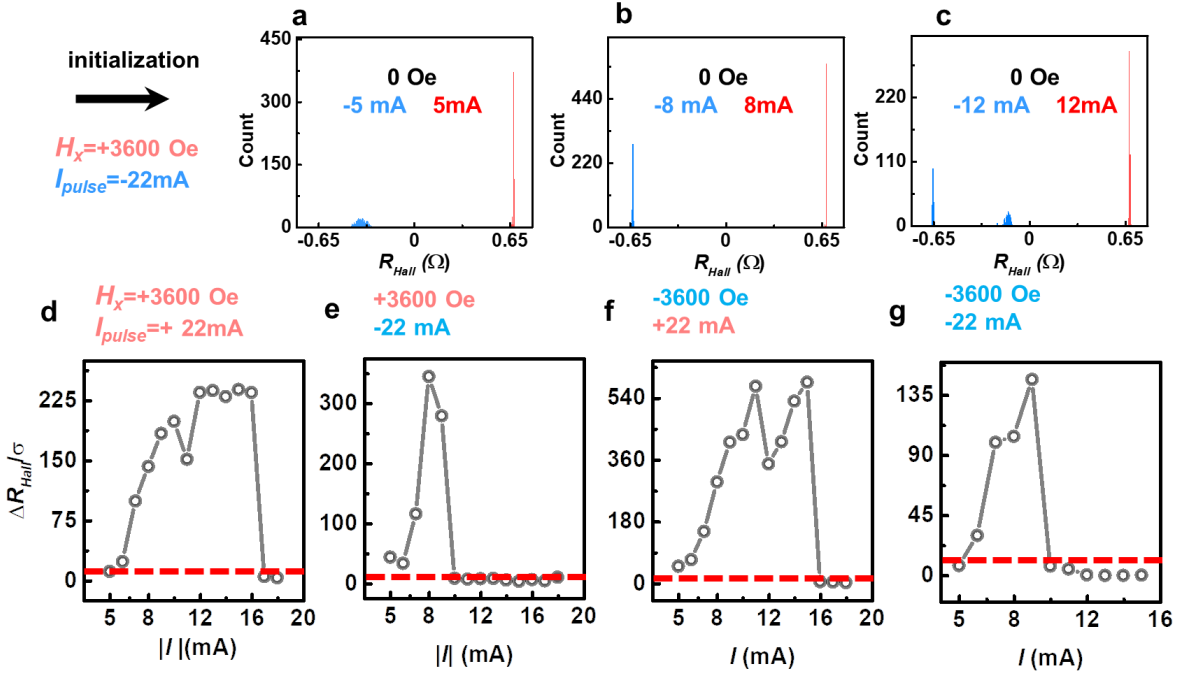

**Figure S3.** (a-c) Field-free deterministic spin–orbit torques switching for initialization of $H_x$=3600 Oe & $I_x$=-22 mA. The alternately positive and negative current pulse sequences of 1000 cycles are applied with duration of 10 ms and amplitude of 5 mA, 8 mA and 12 mA for (a), (b) and (c) respectively. Red line indicates $R_{Hall}$ statistics with positive current pulses, while blue line indicates $R_{Hall}$ statistics with negative current pulses. (d-g) The amplitude of the current pulse $|I|$ versus $\Delta R_{Hall}/\sigma$, which is the average change in resistance from the low- to high-state ($\Delta R_{Hall}$) divided by averaged standard deviation of two states ($\sigma$), for four types of initializations of $H_x$=3600 Oe & $I_x$=22 mA (d), $H_x$=3600 Oe & $I_x$=-22 mA (e), $H_x$=-3600 Oe & $I_x$=22 mA (f) and $H_x$=-3600 Oe & $I_x$=-22 mA (g). Red dashed line in (d)-(g) indicates the line where $\Delta R_{Hall}/\sigma$ equals to 12.

It is extremely important for the rewritable function to have high endurance and wide operation window for applications. To check the endurance and operation window, we perform the endurance measurements under different amplitudes of alternately positive and negative current pulses for 1000 cycles. As shown in Fig. S3a, after initialization of $H_x$=3600 Oe, $I_{pulse}$=-22 mA, a current pulse of -5 mA prefers down-magnetization with $R_{Hall}$=-0.37 Ω, and a current pulse of 5 mA prefers up-magnetization with $R_{Hall}$=0.67 Ω. The ideal case is that both up- and down-magnetization states show the magnitude of 0.67 Ω. The much smaller magnitude at -5 mA indicates an incompletely switching. As shown in Fig. S3b, a current pulse of -8 mA prefers down-magnetization with $R_{Hall}$=-0.64 Ω, and a current pulse of +8 mA prefers up-magnetization with $R_{Hall}$=0.67 Ω. Thus, the added-up amplitude of +8 mA pulse induced magnetization switching is 1.31 Ω, which is comparable to the amplitude of field-induced switching (1.34 Ω). However, further increasing the current amplitude to 12 mA, except for the two main peaks at 0.67 Ω, another resistance level at -0.14 Ω can also be observed, indicating a multi-domain in the cross area. Fig. S3d-g show the $\Delta R/\sigma$ versus current amplitude $|I|$ under four types of initializations, and the operation windows ensuring that $\Delta R/\sigma$ larger than 12 are 10 mA, 4 mA, 10 mA and 4 mA for $H_x$=3600 Oe & $I_x$=22 mA, $H_x$=3600 Oe & $I_x$=-22 mA, $H_x$=-3600 Oe & $I_x$=22 mA and $H_x$=-3600

Oe & $I_x$=-22 mA, respectively. Therefore, we achieved a wide electric operation window of 10 mA for rewritable magnetic domain-wall memory, which meets the requirements for future applications.

**S3. Switching probability of the device from rewritable to read-only function**

The switching probability of the device from rewritable to read-only function was measured after initialization of $H_x$=3600 Oe & $I_x$=22 mA. First, ten cycles of alternately current pulses with amplitude of 9 mA and duration of 10 ms were applied to confirm the realization of rewritable function. Then, a current pulse $I$ was applied to switch the device to read-only function, and followed by a set of pulses (-9mA and +9 mA) to detect whether the switching was successful or not. After each pulse, a small current of 0.1 mA was applied to detect the anomalous Hall resistance. To investigate the switching probability, we counted the number (n) of successful switching to read-only function in 10000 pulses, and took n/10000 as the switching probability. The switching probability for the measured current pulse ranged from 14.5 to 22 mA is shown in Fig. 3d in the main manuscript.

**References and Notes.**